\documentclass[12pt,thmsa]{article}
\usepackage{sw20lart}

\input{tcilatex}
\begin{document}

\author{Krzysztof Ma\'{s}lanka \\
{\small Astronomical Observatory of the Jagiellonian University}\\
{\small \ Orla 171, 30-244 Krak\'{o}w, Poland}}
\title{Series Representation of the Modified Bessel Functions\thanks{%
Cracow Observatory preprint, no. 7/98}}
\date{{\small April 7, 1998 }}
\maketitle

\begin{abstract}
Some power series representations of the modified Bessel functions (McDonald
functions $K_{\alpha }$) are derived using the little known formalism of
fractional derivatives. The resulting summation formulae are believed to be
new.
\end{abstract}

\section{Fractional derivatives}

There are several non-trivial examples in mathematics when some quantity,
originally defined as integer, can radically extend its original range and
assume fractional or even continuous values. The most common example is the
gamma-function of Euler which is a natural, unique generalization of the
ordinary factorial: 
\[
n!\equiv \prod\limits_{i=1}^{n}i=\Gamma \left( n+1\right)
=\int\limits_{0}^{\infty }t^{n}e^{-t}\text{d}t\qquad \left( n>-1\right) 
\]
The same thing may be performed with the order of derivatives which can also
be made fractional. Although useful, fractional derivatives do not however
create any essentially new calculus being rather some sort of particular,
relatively simple, integral transforms.

Following Oldham and Spanier (see e.g. \cite{Oldham}; cf. also \cite
{Vladimiroff}, \cite{Camporesi} and \cite{Maslanka}) we define the
fractional derivative $\partial _{x-a}^{s}\equiv \left( \frac{\partial }{%
\partial x}\right) ^{s}$ by an integral representation known as the
Riemann-Liouville integral. Given a real number $s<0$, define 
\begin{equation}
\partial _{x-a}^{s}f\left( x\right) =\frac{1}{\Gamma \left( -s\right) }%
\int\limits_{a}^{x}\left( x-t\right) ^{-s-1}f\left( t\right) \text{d}t
\label{M1}
\end{equation}
where $a<x$ is a fixed number, referred to as the boundary point. Since $s<0$%
, the integral is convergent, provided that $f$ behaves well. For $s\geq 0$
we define 
\[
\partial _{x-a}^{s}\equiv \partial _{x}^{n}\partial _{x-a}^{s-n} 
\]
where $n$ is a positive integer chosen large enough so that $s-n<0$ in order
to assure convergence of the integral in definition (\ref{M1}). It is not
difficult to show that, as expected, such a definition does not depend on $n$%
. One can further prove that the familiar Leibniz rule for product
differentiation has the form 
\begin{equation}
\partial _{x-a}^{s}\left( fg\right) =\sum\limits_{j=0}^{\infty }\binom{s}{j}%
\partial _{x-a}^{s-j}f~\partial _{x-a}^{j}g  \label{M2}
\end{equation}
where the generalized binomial is 
\[
\binom{s}{j}\equiv \frac{\Gamma \left( 1+s\right) }{j!\Gamma \left(
1+s-j\right) }=\frac{\left( -1\right) ^{j}}{j!}\frac{\Gamma \left(
j-s\right) }{\Gamma \left( -s\right) } 
\]
It can also be shown that the following rule for the arbitrary order
derivative of the power function holds 
\begin{equation}
\partial _{x-a}^{s}\left( x-a\right) ^{p}=\frac{\Gamma \left( p+1\right) }{%
\Gamma \left( p+1-s\right) }\left( x-a\right) ^{p-s}\stackrel{s=n}{%
\rightarrow }\frac{p!}{\left( p-n\right) !}\left( x-a\right) ^{p-n}
\label{M3}
\end{equation}
for any $s$ and $p>-1$. However, the analogous formulae in the case of the
exponential and logarithm functions may, on the first sight, be quite
surprising 
\[
\frac{\partial ^{s}}{\partial x^{s}}\exp \left( \beta x\right) =\beta
^{s}\exp \left( \beta x\right) \frac{\gamma \left( -s,\beta x\right) }{%
\Gamma \left( -s\right) }\stackrel{s=n}{\rightarrow }\beta ^{n}\exp \left(
\beta x\right) 
\]
\[
\frac{\partial ^{s}}{\partial x^{s}}\ln x=\frac{x^{-s}}{\Gamma \left(
1-s\right) }\left[ \ln x-\psi \left( -s\right) -C+\frac{1}{s}\right] 
\stackrel{s=n}{\rightarrow }\frac{\left( -1\right) ^{n-1}\left( n-1\right) !%
}{x^{n}} 
\]
(Here $n$ denotes a positive integer. For brevity we have set the boundary
point $a=0$.)

\section{Some definite integrals as fractional derivatives}

Let us now consider the following definite integrals (cf. \cite{Ryzhik},
formulae 3.471.4 and 3.471.8) 
\begin{equation}
\int\limits_{0}^{x}t^{-2\mu }\left( x-t\right) ^{\mu -1}\exp \left( -\frac{%
\beta }{t}\right) \text{d}t=\frac{\beta ^{\frac{1}{2}-\mu }}{\sqrt{\pi x}}%
\exp \left( -\frac{\beta }{2x}\right) \Gamma \left( \mu \right) K_{\mu -%
\frac{1}{2}}\left( \frac{\beta }{2x}\right)  \label{M4a}
\end{equation}
\begin{equation}
\int\limits_{0}^{x}t^{-2\mu }\left( x^{2}-t^{2}\right) ^{\mu -1}\exp \left( -%
\frac{\beta }{t}\right) \text{d}t=\frac{1}{\sqrt{\pi }}\left( \frac{2}{\beta 
}\right) ^{\mu -\frac{1}{2}}x^{\mu -\frac{3}{2}}\Gamma \left( \mu \right)
K_{\mu -\frac{1}{2}}\left( \frac{\beta }{x}\right)  \label{M4b}
\end{equation}
which are valid for $x>0,\func{Re}\beta >0,\func{Re}\mu >0$. On using (\ref
{M1}) it is evident that both (\ref{M4a}) and (\ref{M4b}) may simply be
interpreted as fractional derivatives of the appropriate functions. Indeed 
\begin{equation}
\frac{\partial ^{s}}{\partial x^{s}}\left[ x^{2s}\exp \left( -\frac{\beta }{x%
}\right) \right] =\frac{\beta ^{s+\frac{1}{2}}}{\sqrt{\pi x}}\exp \left( -%
\frac{\beta }{2x}\right) K_{s+\frac{1}{2}}\left( \frac{\beta }{2x}\right)
\label{M5a}
\end{equation}
\begin{equation}
\frac{\partial ^{s}}{\partial x^{s}}\left[ x^{s-\frac{1}{2}}\exp \left( -%
\frac{\beta }{\sqrt{x}}\right) \right] =\frac{2}{\sqrt{\pi }}\left( \frac{%
\beta }{2}\right) ^{s+\frac{1}{2}}x^{\frac{3}{4}-\frac{s}{2}}K_{s+\frac{1}{2}%
}\left( \frac{\beta }{x}\right)  \label{M5b}
\end{equation}
It is thus possible even to \textit{define }the functions $K_{\nu }$ as
fractional derivatives with zero boundary point.

\section{Polynomials $V_{k}$}

In order to evaluate the left hand side of (\ref{M5a}) and (\ref{M5b}) let
us introduce 
\begin{equation}
V_{k}^{\left( \alpha \right) }\left( \beta x^{\alpha }\right) \equiv
x^{k}\exp \left( \beta x^{\alpha }\right) \frac{\partial ^{k}}{\partial x^{k}%
}\exp \left( -\beta x^{\alpha }\right)  \label{M6}
\end{equation}
It is easy to prove that, contrary to their appearance, $V_{k}$ are just
polynomials 
\[
V_{k}^{\left( \alpha \right) }\left( z\right)
=\sum\limits_{j=0}^{k}A_{kj}^{\left( \alpha \right) }z^{j} 
\]
with coefficients given by 
\[
A_{kj}^{\left( \alpha \right) }=\left( -1\right) ^{k}\sum\limits_{i=0}^{j}%
\frac{\left( -1\right) ^{i}}{i!\left( j-i\right) !}\frac{\Gamma \left(
k-\alpha i\right) }{\Gamma \left( -\alpha i\right) } 
\]
In particular, for $\alpha =-1$ we have 
\[
A_{kj}^{\left( -1\right) }=\left( -1\right) ^{k}\sum\limits_{i=0}^{j}\frac{%
\left( -1\right) ^{i}}{i!\left( j-i\right) !}\frac{\Gamma \left( k+i\right) 
}{\Gamma \left( i\right) }=\frac{\left( -1\right) ^{k+j}}{\left( k-j\right) !%
}\frac{k!\left( k-1\right) !}{j!\left( j-1\right) !} 
\]
(The last equality may be rigorously proved using elementary methods
presented in e.g. \cite{Graham}; \textit{Mathematica} 3.0 effectively
simplifies such sums.) We shall also need fractional derivatives of the
following expressions 
\[
x^{\nu }\exp \left( -\beta x^{\alpha }\right) 
\]
with $\alpha ,\nu $ real. Using Leibniz rule (\ref{M2}) and the property (%
\ref{M3}) we have 
\begin{equation}
\frac{\partial ^{s}}{\partial x^{s}}\left[ x^{\nu }\exp \left( -\beta
x^{\alpha }\right) \right] =x^{\nu -s}\frac{\Gamma \left( \nu +1\right) }{%
\Gamma \left( -s\right) }\exp \left( -\beta x^{\alpha }\right)
\sum\limits_{k=0}^{\infty }\frac{\left( -1\right) ^{k}}{k!}\frac{\Gamma
\left( k-s\right) }{\Gamma \left( k-s+\nu +1\right) }V_{k}^{\left( \alpha
\right) }  \label{M7}
\end{equation}

\section{Series expansion of the Modified Bessel functions}

Inserting (\ref{M7}) into (\ref{M5a}) or (\ref{M5b}) we can get a general
expansion of the modified Bessel functions (called also McDonald functions;
cf. e.g. \cite{Abramowitz}) holding for any positive $\alpha $%
\begin{equation}
K_{s}\left( z\right) =\frac{\sqrt{\pi }}{\left( 2z\right) ^{s}}\exp \left(
-z\right) \frac{\Gamma \left( 2s\right) }{\Gamma \left( \frac{1}{2}-s\right) 
}\sum\limits_{k=0}^{\infty }\frac{\left( -1\right) ^{k}}{k!}\frac{\Gamma
\left( k+\frac{1}{2}-s\right) }{\Gamma \left( k+\frac{1}{2}+s\right) }%
V_{k}^{\left( -1\right) }\left( 2z\right)  \label{M9}
\end{equation}
Substituting polynomials (\ref{M6}) into (\ref{M9}) we further get, after
some manipulations 
\[
K_{s}\left( z\right) =2^{s-1}\Gamma \left( s\right) z^{-s}\exp \left(
-z\right) \left[ 1+\sum\limits_{k=1}^{\infty }\frac{\left( \frac{1}{2}%
-s\right) _{k}}{\left( \frac{1}{2}+s\right) _{k}}\sum\limits_{j=1}^{k}%
\QOVERD( ) {k-1}{j-1}\frac{\left( -2z\right) ^{j}}{j!}\right] 
\]
where 
\[
\left( a\right) _{k}\equiv \frac{\Gamma \left( k+a\right) }{\Gamma \left(
a\right) } 
\]
denotes the Pochhammer symbol. The same method applied to (\ref{M5b}) gives 
\begin{equation}
K_{s}\left( z\right) =2^{s-1}\sqrt{\pi }\frac{\Gamma \left( 2s\right) }{%
\Gamma \left( \frac{1}{2}-s\right) }z^{-s}\exp \left( -z\right)
\sum\limits_{k=0}^{\infty }\frac{\left( -1\right) ^{k}}{k!}\frac{\Gamma
\left( k+\frac{1}{2}-s\right) }{\Gamma \left( k+\frac{1}{2}+s\right) }%
V_{k}^{\left( -1/2\right) }\left( z\right)  \label{M10}
\end{equation}
which is yet another power series convergent to the McDonald function. The
familiar ratio of Pochhammer symbols in (\ref{M9}) and (\ref{M10}) allows us
to call these expansions 'hypergeometric-like' since they formally resemble
the well-known hypergeometric Kummer function $_{1}F_{1}$.


\begin{thebibliography}{9}
\bibitem{Oldham}  K. B. Oldham, J. Spanier, \textit{The Fractional Calculus}%
, Academic Press, New York, 1974

\bibitem{Vladimiroff}  V. S. Vladimiroff, \textit{Obobshchennye funkcii v
matematicheskoj fizikie}, Nauka, Moskva, 1979 (in Russian)

\bibitem{Camporesi}  R. Camporesi, \textit{Harmonic Analysis and Propagators
on Homogeneous Spaces}, Physics Reports 196 no. 1\&2, 1990

\bibitem{Maslanka}  K. Ma\'{s}lanka, \textit{Efekty kwantowe w
wielowymiarowych modelach kosmologicznych}, Ph. D. thesis, 1986,
(unpublished, in Polish)

\bibitem{Ryzhik}  I. S. Ryzhik, I. M. Gradstein, \textit{Tablicy
intiegralov, sum, riadow i proizvedenij}, Nauka, Moskva, 1971, (in Russian)

\bibitem{Abramowitz}  M. Abramowitz, I. A. Stegun, eds., \textit{Handbook of
Mathematical Functions}, U.S. National Bureau of Standards, Dover, New York,
1965

\bibitem{Graham}  R. L. Graham, D. E. Knuth, O. Patashnik, \textit{Concrete
Mathematics. A Foundation for Computer Science}, Addison-Wesley, 1994
\end{thebibliography}
\end{document}